\documentclass[pra,twocolumn,superscriptaddress,showpacs,preprintnumbers,amsmath,amssymb]{revtex4}
\usepackage{mathrsfs}
\usepackage{bbm}
\usepackage{amsfonts}
\usepackage{tipa}

\usepackage{epsfig,graphicx}
\usepackage{amstext}
\usepackage{amsmath}
\usepackage{graphicx}
\usepackage{times}

\begin{document}


\title{Role of weak measurements on states ordering and monogamy of quantum correlation}

\author{Ming-Liang Hu}
\email{mingliang0301@163.com}
\affiliation{School of Science, Xi'an Jiaotong University,
             Xi'an 710049, China}
\affiliation{School of Science, Xi'an University of Posts and
             Telecommunications, Xi'an 710121, China}
\author{Heng Fan}
\affiliation{Beijing National Laboratory for Condensed Matter Physics,
             Institute of Physics, Chinese Academy of Sciences, Beijing
             100190, China}
\author{Dong-Ping Tian}
\affiliation{School of Science, Xi'an Jiaotong University,
             Xi'an 710049, China}

\begin{abstract}
The information-theoretic definition of quantum correlation, e.g.,
quantum discord, is measurement dependent. By considering the more
general quantum measurements, weak measurements, which include the
projective measurement as a limiting case, we show that while weak
measurements can enable one to capture more quantumness of
correlation in a state, it can also induce other counterintuitive
quantum effects. Specifically, we show that the general measurements
with different strengths can impose different orderings for quantum
correlations of some states. It can also modify the monogamous
character for certain classes of states as well which may diminish
the usefulness of quantum correlation as a resource in some
protocols. In this sense, we say that the weak measurements play a
dual role in defining quantum correlation.
\end{abstract}

\pacs{03.65.Ud, 03.65.Ta, 03.67.Mn
\\Key Words: Quantum discord; Weak measurement; Monogamy
}

\maketitle

\section{Introduction}\label{sec:1}

Quantum correlation plays a crucial role in many quantum computation
and information processing tasks \cite{Horodecki}. Quantum discord
(QD) \cite{Ollivier,Henderson,Dakic,Modi}, which goes beyond the
traditional measure of quantum correlation, i.e., quantum
entanglement, is proposed to be responsible for the power of a mixed
state quantum algorithm with vanishing or negligible entanglement
\cite{Datta}. It has potential applications in detecting critical
points of quantum phase transitions even at finite temperatures
\cite{qpt,qpt1,qpt2}. QD is also found to be the necessary resource
for remote state preparation \cite{rsp}, quantum state discrimination
\cite{qsd,qsd1}, and quantum locking \cite{qlock,qlock1}. A connection
between QD consumption and the quantum advantage for encoding
information has been identified as well \cite{Gumile}. These findings
have prompted a huge surge of interest in understanding QD from different
perspectives, such as its operational interpretation via quantum
state merging \cite{Madhok,Cavalcanti} and teleportation fidelity
\cite{Adhikari}, the generation of QD via local operations
\cite{local1,local2,local3,local4,local5}, the discording power
of nonlocal unitary gates \cite{dpower}, and other related issues
of QD \cite{ad1,ad2,ad3,ad4,ad5}; see a recent review paper
\cite{Modirmp} for more results.

Due to the fundamental significance and potential applications,
various measures of QD \cite{Ollivier,Henderson,Dakic,Modi}, as
well as other related measures of quantum correlations
\cite{Giorgi,mea1,mea2,mea3,mea4,mea5}, have been introduced. The general
positive operator valued measurements (POVMs) are proposed in the
original definitions of these measures. On the other hand, in view
of the generally negligible improvement by doing minimization over
full POVMs \cite{Hamieh,Galve}, measures of quantum correlation are
usually evaluated by restricting to only projective measurements. The
process of projective measurements are performed by constructing a
set of orthogonal projectors in the Hilbert space of a Hermitian
operator $\mathcal {A}$, and the possible outcomes of the
measurements are given by the spectra of $\mathcal {A}$. This
process usually induces strong perturbations to the measured system,
and possibly constrains one's ability to extract as much quantum
correlations as possible.

Weak measurement can provides new insights into the study of some
fundamental problems of quantum mechanics and has already been
realized experimentally \cite{add1,add2,add3}. Also it can be used
for signal amplification practically and for state tomography
\cite{WuSJ}. Particularly, these measurement processes are universal
in that any generalized measurements can be decomposed into a
sequence of weak measurements \cite{weak2}. Since its fundamental
role in quantum theory and practical applications, it is natural to
consider the quantumness of correlations by weak measurement. The
weak measurement which can be implemented by coupling the system to
the measurement apparatus weakly and generally have small influence
on a quantum state due to the partial collapsing of the measured
wavefunction \cite{weak1,weak2,weak3,weak4,weak5}. This differs it
from projective measurement performed in standard QD. The quantum
correlation based on weak measurements is proposed as
super quantum discord \cite{Singh}. This new quantifier has been
shown to play a potential role in the protocol of optimal assisted
state discrimination where entanglement is totally not necessary
\cite{libo}, and it has also stimulated other related definition of
quantum correlations \cite{feism}.

The super discord is always larger than the normal discord defined
by the strong (projective) measurements \cite{Singh}, and this may
be regarded as a figure of merit by using the weak measurements in
characterizing quantumness of correlations in a state. But just as
every coin has two sides, here we will show that the use of weak
measurements in defining quantum correlations can also induce other
counterintuitive effects. As explicit examples, we will show that
the super discord captured by the weak measurements with different
strengths can impose different orderings of quantum states. This
phenomenon is very different from those of the states ordering
obtained in the literature \cite{order1,order2,order3,order4,order5},
which are easy to understand as they are induced by different
correlation measures, e.g., the entropic measure of discord
\cite{Ollivier} and the geometric measure of discord \cite{Dakic}.
Moreover, we will also show that the super discord can change the
monogamy nature for certain classes of states \cite{Giorgigl,Prabhu,
Sudha,Streltsov,Bera,Braga,Fanchini,Ren}. Detailed examples show
that this change presents in a wide class of quantum states, and
therefore may result in failure of certain quantum tasks, such as
the protocol that distinguishes the generalized
Greenberger-Horne-Zeilinger (GHZ) states from the generalized {\it W}
states by using the monogamy conditions of QD \cite{Prabhu}.

\section{Definition of super discord}\label{sec:2}
In this section, we will introduce the concept of super discord
\cite{Singh}. Its definition is somewhat similar as that of the
normal discord introduced by Ollivier and Zurek \cite{Ollivier}. The
only difference is that the original projective operators are
replaced by the weak measurement operators of the following form
\cite{weak2}
\begin{eqnarray}\label{eq1}
 \mathcal {P}_{+}(x)&=&\sqrt{\frac{1-\tanh x}{2}}\Pi_0
                       +\sqrt{\frac{1+\tanh x}{2}}\Pi_1,\nonumber\\
 \mathcal {P}_{-}(x)&=&\sqrt{\frac{1+\tanh x}{2}}\Pi_0
                       +\sqrt{\frac{1-\tanh x}{2}}\Pi_1,
\end{eqnarray}
where the strength of the measurement process is parameterized by a
parameter $x\in \mathbb{R}$. $\Pi_0$ and $\Pi_1$ are the orthogonal
projectors summing to the identity, and therefore $\mathcal
{P}_{+}^2+\mathcal {P}_{-}^2=I$.  Along this line of measurement
formalism, one can then obtain the nonselective postmeasurement
state as
\begin{eqnarray}\label{eq2}
 \rho_{A|\mathcal {P}_{\pm}^B}=\frac{{\rm Tr}_B[(I\otimes \mathcal {P}_{\pm}^B)\rho_{AB}
                               (I\otimes \mathcal {P}_{\pm}^B)]}{p_\pm},
\end{eqnarray}
after the weak measurements being performed on party $B$, and $p_\pm
={\rm Tr}[(I\otimes \mathcal {P}_\pm^B)\rho_{AB}(I\otimes \mathcal
{P}_\pm^B)]$ is the probability distribution for the measurement
outcomes. Then the super discord is defined as
\begin{eqnarray}\label{eq3}
 D_w^{\leftarrow}(\rho_{AB})=\min_{\{\Pi_k^B\}}S_w(A|\{\mathcal {P}_\pm^B\})-S(A|B),
\end{eqnarray}
where the minimization is taken over the complete set of the
projection-valued measurements $\{\Pi_k^B\}$. The conditional von
Neumann entropy for the premeasurement state $\rho_{AB}$ is denoted
by $S(A|B)=S(\rho_{AB})-S(\rho_B)$, while the averaged conditional
von Neumann entropy for the postmeasurement state is denoted by
\begin{eqnarray}\label{eq4}
 S_w(A|\{\mathcal {P}_\pm^B\})=p_{+}S(\rho_{A|\mathcal {P}_{+}^B})
                               +p_{-}S(\rho_{A|\mathcal {P}_{-}^B}),
\end{eqnarray}
with $\rho_{B(A)}={\rm Tr}_{A(B)}\rho_{AB}$ being the reduced
density operator of $\rho_{AB}$, and $S(\rho)=-{\rm
Tr}(\rho\log_2\rho)$ represents the von Neumann entropy \cite{Nielsen}.

\section{States ordering with super discord}\label{sec:3}
Given two quantum correlation measures, $\mathcal {Q}$ and $\mathcal
{R}$, they are said to give the unique states ordering if and only
if the following condition
\begin{eqnarray}\label{eq5}
 \mathcal {Q}(\rho_1)\geqslant \mathcal {Q}(\rho_2)\Longleftrightarrow
                     \mathcal {R}(\rho_1)\geqslant \mathcal {R}(\rho_2)
\end{eqnarray}
is satisfied for arbitrary $\rho_1$ and $\rho_2$.

For the entanglement measures of the {\it concurrence}
\cite{concurrence} and the {\it negativity} \cite{negativity}, or
quantum correlation measures of the {\it entropic discord}
\cite{Ollivier} and the {\it geometric discord} \cite{Dakic}, it is
a well accepted fact that the condition presented in Eq. \eqref{eq5}
can be violated by certain two-qubit mixed states, see, for example,
Refs. \cite{order1,order2,order3,order4}. For higher dimensional
system, entanglement quantified by R\'{e}nyi entropies may also
violate this condition \cite{order5}. But these violations are
conceptually easy to understand as they are induced by the
correlation measures defined from different perspectives. Here, as
an unexpected result, we will show that even under the framework of
weak measurements with different strengths \cite{weak2}, the
resulting super discords similarly do not necessarily imply the same
orderings of quantum states.

In the following, we illustrate the above argument through an
explicit example. We consider a family of two-qubit states with
maximally mixed marginals
\begin{eqnarray}\label{eq6}
 \rho_{AB}=\frac{1}{4}(I\otimes I+\sum_{i=1}^{3} c_i \sigma_i\otimes\sigma_i),
\end{eqnarray}
where $I$ denotes the $2\times 2$ identity operator, and
$\sigma_{1,2,3}$ are the usual Pauli operators. Physical $\rho_{AB}$
are those with $(c_1,c_2,c_3)$ being confined to the tetrahedron
with vertices $(-1, -1, -1)$, $(-1,1,1)$, $(1, -1,1)$, and $(1,1,
-1)$ \cite{tetrahedron}. These states are usually termed as the
Bell-diagonal states as they can be decomposed into linear
combinations of the four Bell states. The super discord for
$\rho_{AB}$ is calculated as \cite{feism}
\begin{eqnarray}\label{eq7}
 D_w^{\leftarrow}(\rho_{AB})&=&\frac{1-c_1-c_2-c_3}{4}\log_2(1-c_1-c_2-c_3)\nonumber\\
               &&+\frac{1-c_1+c_2+c_3}{4}\log_2(1-c_1+c_2+c_3)\nonumber\\
               &&+\frac{1+c_1-c_2+c_3}{4}\log_2(1+c_1-c_2+c_3)\nonumber\\
               &&+\frac{1+c_1+c_2-c_3}{4}\log_2(1+c_1+c_2-c_3)\nonumber\\
               &&-\frac{1-c\tanh x}{2}\log_2 (1-c\tanh x)\nonumber\\
               &&-\frac{1+c\tanh x}{2}\log_2 (1+c\tanh x),
\end{eqnarray}
where $c=\max\{|c_1|,|c_2|,|c_3|\}$.

\begin{figure}
\centering
\resizebox{0.45\textwidth}{!}{%
\includegraphics{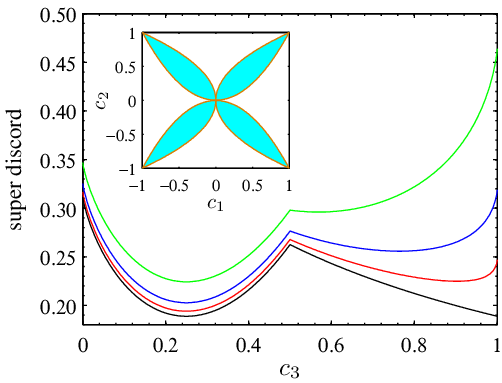}}
\caption{(Color online) Super discord versus $c_3$ for the
Bell-diagonal
         states of Eq. \eqref{eq6} with $c_1=0.5$ and $c_2=-0.5$. The
         green, blue, red, and black lines (from top to bottom) are plotted
         with $x=1.5$, $2.0$, $2.5$, and $\infty$ (i.e., the normal discord),
         respectively. The cyan (gray) shaded region in the inset represents valid
         $(c_1,c_2)$ for which the Bell-diagonal states obtained by varying
         $c_3$ have different orderings induced by the super discord with
         different measurement strengths.} \label{fig:1}
\end{figure}

In Fig. \ref{fig:1}, we present an exemplified plot of the super
discord as functions of $c_3$ for the Bell-diagonal states of Eq.
\eqref{eq6} with $c_1=0.5$ and $c_2=-0.5$. The four curves from top
to bottom are obtained by choosing the controlling parameters as
$x=1.5$, $2.0$, $2.5$, and $\infty$ (corresponds to the normal
discord), respectively, from which one can observe that there are
states having different orderings induced by the super discord with
the weak measurements of different strengths. This counterintuitive
phenomenon can be further confirmed by the cyan (gray) shaded region
shown in the inset of Fig. \ref{fig:1}, which stands for the valid
$(c_1,c_2)$ for which $\rho_{AB}$ of Eq. \eqref{eq6} with different
$c_3$ can have different states ordering. It provides an intriguing
perspective of the super discord in that it implies the quantum
correlation in a state is not only measurement-method-dependent \cite{Espagnat,long} but
is also strongly dependent on the internal structures of the related
measurements.

\begin{figure}
\centering
\resizebox{0.45\textwidth}{!}{%
\includegraphics{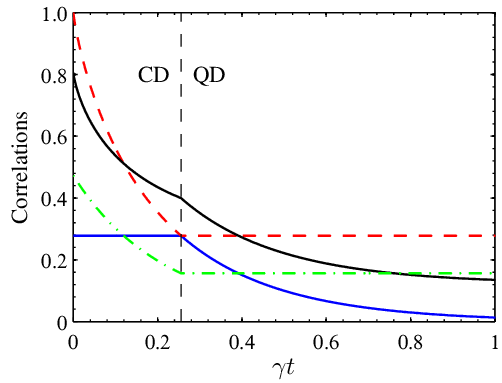}}
\caption{(Color online) Normal discord [blue (gray) solid line],
         classical correlation (red dashed line), as well as the super
         discord (black solid line) and super classical correlation (green
         dash-dot line) obtained with $x=1.0$ versus $\gamma t$ under phase
         damping channel for the initial Bell-diagonal state of Eq.
         \eqref{eq6} with $c_1(0)=1$, $c_2(0)=-0.6$, and $c_3(0)=0.6$. CD and
         QD denote, respectively, the dynamical regimes of classical
         decoherence and quantum decoherence with respect to the projective
         measurement.} \label{fig:2}
\end{figure}

The lack of the unique states ordering may results in completely
different dynamical behaviors of quantum correlations with respect
to super discord. As an example, we consider the case of two qubits
being prepared initially in the Bell-diagonal state of Eq.
\eqref{eq6}, and subject to the same phase damping channels \cite{Nielsen}, the
process of which preserves the Bell-diagonal form of $\rho_{AB}$,
with however the time dependence of the three parameters being given
by $c_1(t)=c_1(0)\exp(-2\gamma t)$, $c_2(t)=c_2(0)\exp(-2\gamma t)$,
and $c_3(t)=c_3(0)$, where $\gamma$ is the phase damping rate.

For this kind of local dissipative channel, it has been found that
there are sudden transition from dynamical regimes of the classical
decoherence (CD) to the quantum decoherence (QD) when considering
the normal discord defined via the projective measurements
\cite{Mazzola}, see, for example, the blue (gray) solid and red
dashed lines shown in Fig. \ref{fig:2}. When considering quantum and
classical correlations based on the paradigm of weak measurements,
however, the original two distinct regimes disappear. As can be seen
from the black solid line shown in Fig. \ref{fig:2}, the super
discord decays with increasing $\gamma t$ even in the CD regime,
which is in sharp contrast to that of the normal discord who is
constant in time during the same regime.

Inspired by the connection between the normal discord and classical
correlation \cite{Ollivier}, we further define the super classical
correlation as
\begin{equation}\label{eq8}
 C_w^{\leftarrow}(\rho_{AB})=I(\rho_{AB})-D_w^{\leftarrow}(\rho_{AB}),
\end{equation}
with $I(\rho_{AB})=S(\rho_A)+S(\rho_B)-S(\rho_{AB})$ being the
quantum mutual information \cite{Nielsen}. Then with the same
parameters as those for the super discord in Fig. \ref{fig:2}, we
presented dynamics of $C_w^{\leftarrow}(\rho_{AB})$ as the green
dash-dot line in the same figure, from which one can note that it
displays qualitatively similar behaviors as that for the normal
classical correlation, i.e., it decays with time in the CD regime,
and remains constant in the QD regime. We thus see that although the
weak measurements can change the dynamical behaviors of super discord
in the CD regime, it has no influence on the qualitative behaviors
of the classical correlation. In fact, the unique ordering of states
with the super classical correlations is universal for the class
of Bell-diagonal states of Eq. \eqref{eq6}, for which we always have
\begin{eqnarray}\label{eq9}
 C_w^{\leftarrow}(\rho_{AB})&=&\frac{1-c\tanh x}{2}\log_2 (1-c\tanh x)\nonumber\\
               &&+\frac{1+c\tanh x}{2}\log_2 (1+c\tanh x),
\end{eqnarray}
which can be shown to be a monotonic increasing function of $c$ for
arbitrary $x$, and therefore it always impose the same ordering for
the Bell-diagonal states. But it should be note that the above
argument does not hold for general case, as there are $\rho_1$ and
$\rho_2$ such that the unique ordering condition is violated.

\section{Monogamy of super discord}\label{sec:4}
We now turn to discuss the role weak measurements played in
exploring monogamous character of quantum correlations. Due to the
asymmetry of the super discord \cite{Singh}, there are two possible
lines of research on this problem, which can be illustrated
explicitly through the following two monogamy inequalities
\cite{Giorgigl,Prabhu,Sudha,Streltsov,Bera,Braga,Fanchini,Ren}
\begin{eqnarray}\label{eq10}
 &&D_w^{\leftarrow}(\rho_{A:BC}) \geqslant
         D_w^{\leftarrow}(\rho_{AB})+D_w^{\leftarrow}(\rho_{AC}),\nonumber\\
 &&D_w^{\rightarrow}(\rho_{A:BC}) \geqslant
         D_w^{\rightarrow}(\rho_{AB})+D_w^{\rightarrow}(\rho_{AC}),
\end{eqnarray}
where the first one is formulated with the measurements being
performed on different subsystems $B$, $C$ and $BC$ of $\rho_{ABC}$
\cite {Giorgigl,Prabhu,Sudha,Streltsov,Bera,Braga,Fanchini}, and the
second one is formulated with the measurements being performed on
the same subsystem $A$ \cite{Ren}. For convenience of later
presentation, we further define $\Delta D_w^{\alpha}
=D_w^{\alpha}(\rho_{A:BC})-D_w^{\alpha}(\rho_{AB})
-D_w^{\alpha}(\rho_{AC})$ ($\alpha\in\{\leftarrow,\rightarrow\}$) as
the related discord monogamy score \cite{Bera}.

As the super discord is an extension of the normal discord, and the
normal discord has been found to be monogamous for certain classes
of quantum states (e.g., the generalized GHZ-class state)
\cite{Prabhu}, it is natural to ask whether this monogamy nature is
universal for the super discord with arbitrary measurement
strengths, or whether the super discord still respect monogamy for
these states?

\begin{figure}
\centering
\resizebox{0.45\textwidth}{!}{%
\includegraphics{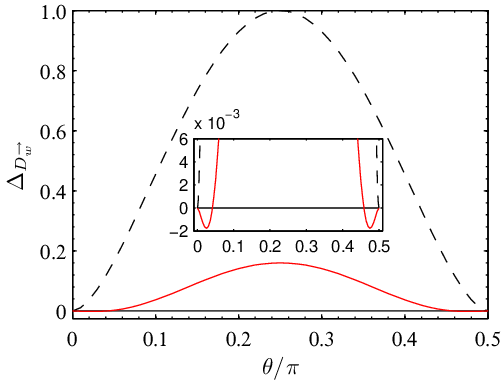}}
\caption{(Color online) Dependence of the discord monogamy score
         $\Delta D_w^{\rightarrow}$ on $\theta/\pi$ for $|\Psi\rangle_{\rm
         GHZ}$. The black dashed and the red solid lines are obtained with
         $x\rightarrow\infty$ and $x=0.5$, respectively. The inset is plotted
         for better visualizing violation of the monogamy condition during
         the small and large $\theta/\pi$ regions.} \label{fig:3}
\end{figure}

Here, we will show through some explicit examples the answer to this
question is indeed state-dependent, that is to say, these states can
be monogamous as well as polygamous with respect to super discord.
Our first exemplification is that of the generalized GHZ states
\begin{eqnarray}\label{eq11}
 |\Psi\rangle_{\rm GHZ}=\cos\theta|000\rangle+\sin\theta|111\rangle,
\end{eqnarray}
which is known to be monogamous with respect to the normal discord
with infinite measurement strength $x$ \cite{Prabhu}. When
considering the super discord defined with finite $x$, the two
monogamy conditions in Eq. \eqref{eq10} are in fact equivalent due
to the exchange symmetry of $|\Psi\rangle_{\rm GHZ}$. In Fig.
\ref{fig:3} we plotted the related monogamy score $\Delta
D_w^{\rightarrow}$ against $\theta/\pi$ with the measurement
strengths $x\rightarrow\infty$ (corresponds to the normal discord)
and $x=0.5$, respectively. The states are monogamous whenever
$\Delta D_w^{\rightarrow}$ take positive values. This figure shows
evident transitions from observation to violation of monogamy for
the super discord. More specifically, the super discord does not
respect monogamy during the small and large regions of $\theta/\pi$.
This result is interesting as it implies that the monogamy property
of discord, even for those defined under the same formalism of
measurements, is not only state-dependent but is also determined by
the intrinsic properties, e.g., the measurement strengths, of the
related measurements.

\begin{figure}
\centering
\resizebox{0.45\textwidth}{!}{%
\includegraphics{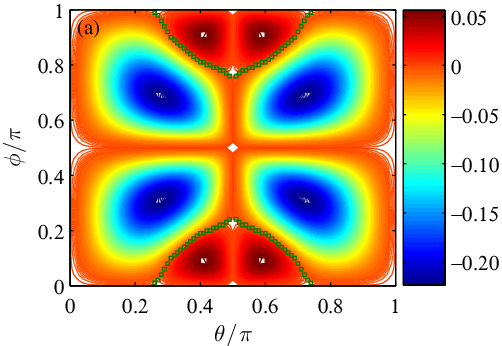}}
\centering
\resizebox{0.45\textwidth}{!}{%
\includegraphics{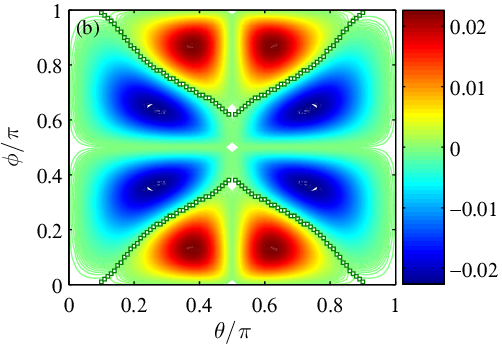}}
\caption{(Color online) Contour plots of the discord monogamy score
         $\Delta D_w^{\rightarrow}$ against $\theta/\pi$ and $\phi/\pi$ for
         $|\Psi\rangle_{W}$ with $x\rightarrow\infty$ (a) and $x=0.5$
         (b), respectively. The regions up and below the green hollow squares
         represent valid $|\Psi\rangle_{W}$ with monogamy.} \label{fig:4}
\end{figure}

As another example, we consider the generalized {\it W}-class states
\begin{eqnarray}\label{eq12}
 |\Psi\rangle_{W}=\sin\theta\cos\phi|011\rangle
                  +\sin\theta\sin\phi|101\rangle
                  +\cos\theta|110\rangle,\nonumber\\
\end{eqnarray}
for which the two monogamy inequalities in Eq. \eqref{eq10} are no
longer equivalent, and when evaluated via the normal discord, the
first one is always violated \cite{Prabhu}, while the second one
may be satisfied or violated \cite{Ren}.

When considering quantum correlations captured by the super discord,
our results revealed that the first monogamy condition in Eq.
\eqref{eq10} still remains violated for the weak measurements of
arbitrary strengths. But if we consider the second condition, the
case will be very different. As can be seen from the contour plots
shown in Fig. \ref{fig:4}, the monogamy property turns out to be
state-dependent, and in contrast to that for the generalized
GHZ-class states, here the $(\theta,\phi)$ regions for monogamy are
enlarged with finite measurement strength.

From a practical point of view, the monogamous nature of the normal
discord can be used to distinguish two stochastic local operations
and classical communication (SLOCC) inequivalent classes of
tripartite states, i.e., the generalized GHZ and {\it W} classes
\cite{Prabhu}. But when evaluated via the super discord, the above
results revealed that this application will does not applicable, as
the GHZ-class states can also be monogamous as well as polygamous
for this case (see, Fig. \ref{fig:3}). This exhibits another
perspective of the weak measurements in defining quantum
correlations, which is beyond our expectation.

\begin{figure}
\centering
\resizebox{0.45\textwidth}{!}{%
\includegraphics{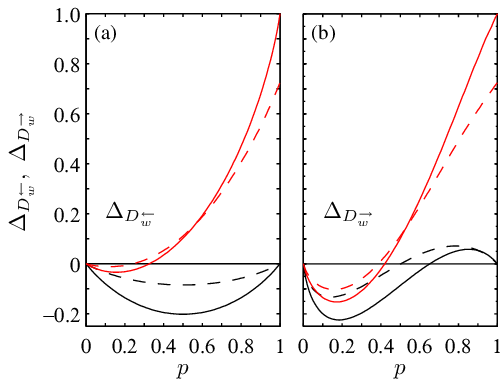}}
\caption{(Color online) Dependence of the discord monogamy score
         $\Delta D_w^{\leftarrow}$ and $\Delta D_w^{\rightarrow}$ on $p$ for
         $|\psi(p,\epsilon)\rangle$. The black and red (gray) solid lines are
         plotted for the normal discord with $\epsilon=1$ and $\epsilon=0.5$,
         respectively; while the black and red (gray) dashed lines are
         plotted for the super discord with $x=1.5$, and $\epsilon=1$ and
         $\epsilon=0.5$, respectively.} \label{fig:5}
\end{figure}

Finally, we remark that the change to the monogamy nature of quantum
states with respect to the super discord can also happen when the
weak measurements are performed on different subsystems. See, for
example, the results displayed in Fig. \ref{fig:5}, which is plotted
for the tripartite pure states $|\psi(p,\epsilon)\rangle
=\sqrt{p\epsilon}|000\rangle +\sqrt{p(1-\epsilon)}|111\rangle
+\sqrt{(1-p)/2}(|101\rangle+|110\rangle)$.

\section{Conclusion}\label{sec:5}
Weak measurements are important complementary to the standard
measurement in quantum theory. By reconsidering the processes of
weak measurements with different strengths, we showed that while
being able to capture more quantum correlation, it can also induce
other counterintuitive effects meanwhile. As the first
exemplification, we showed that the weak measurements with different
strengths can impose different orderings of quantum states. This
effect is very different from those observed with different quantum
correlation measures, e.g., the entropic and the geometric measures
of the normal discord, and it may results in unexpected dynamical
behaviors of quantum correlations. Moreover, we have also showed
that the monogamous nature of the normal discord for certain classes
of quantum states can be changed by the weak-measurement-defined
super discord, and this change can even invalidate the feasibility
of some quantum tasks, such as the detection of two
SLOCC-inequivalent classes of tripartite states based on monogamy
\cite{Prabhu}. In view of these facts, we then conclude that the
weak measurements play a dual role in defining quantum correlations.

On the other hand, since different physical systems may naturally
interact strongly or weakly with probing systems, the full
description of measurement-dependent quantum correlation may be
complete only when weak measurement with adjustable strengths are
considered. This may also provide a full quantification of quantum
correlation restricted experimentally to some specified quantum
systems. In particular, in case the quantum correlation based on
weak measurements may enhance or diminish its usefulness in some
protocols, a complete view of super quantum discord is necessary and
may shed light on our understanding of other quantum
characteristics.

\section*{ACKNOWLEDGMENTS}
This work was supported by NSFC (11205121, 10974247, 11175248), the
``973'' program (2010CB922904), NSF of Shaanxi Province
(2010JM1011), and the Scientific Research Program of the Education
Department of Shaanxi Provincial Government (12JK0986).

\newcommand{\PRL}{Phys. Rev. Lett. }
\newcommand{\RMP}{Rev. Mod. Phys. }
\newcommand{\PRA}{Phys. Rev. A }
\newcommand{\PRB}{Phys. Rev. B }
\newcommand{\NJP}{New J. Phys. }
\newcommand{\JPA}{J. Phys. A }
\newcommand{\JPB}{J. Phys. B }
\newcommand{\PLA}{Phys. Lett. A }
\newcommand{\NP}{Nat. Phys. }
\newcommand{\NC}{Nat. Commun. }
%

%

\end{document}